\documentclass{mn2e}
\usepackage{epsfig}

\title[Shocks in the Atmosphere of BW\,Vul]{Spectral Response of
  the 
Pulsationally-Induced Shocks in the Atmosphere of BW\,Vulpeculae}

\author[M. A. Smith and C. S. Jeffery]{
  Myron  A. Smith$^{1}$\thanks{E-mail:msmith@stsci.edu} and  
  C. Simon Jeffery$^{2}$\thanks{E-mail:csj@star.arm.ac.uk}\\
$^{1}$ Computer Sciences Corporation/STScI, 3700 San Martin Drive, 
        Baltimore, MD 21218 \\
$^{2}$ Armagh Observatory, College Hill, Armagh BT61 9DG, N. Ireland }

\begin{document}

\date{Accepted \ldots. Received \ldots}

\pagerange{\pageref{firstpage}--\pageref{lastpage}} \pubyear{2002}

\maketitle

\label{firstpage}

\begin{abstract}

  BW\,Vul is remarkable for exciting an extremely strong radial pulsation 
mode. This instability grows in its outer envelope and forms 
visible shock features in the continuum flux and spectral line profiles 
at two phases separated by 0.8 cycles. Material propelled upwards 
energetically in the 
atmosphere from the shock returns to the lower photosphere where it 
creates a second shock just before the start of the next cycle. 
We have obtained three nights of echelle data for this star over about
five pulsation cycles (P = 0.201 days) in order to evaluate the effects of 
atmospheric shocks on a number of important red lines in the spectrum. 
These lines include He\,I $\lambda$5875 and $\lambda$6678, C\,II 
$\lambda\lambda$6578-83 doublet, and other moderate 
(e.g., Si\,II $\lambda$6371) 
and high excitation (Si\,III $\lambda$5737) lines. We have added to these data
37 archival IUE/SWP echelle spectra obtained in 1994. We have investigated 
the equivalent widths and shapes of the optical lines for evidence of 
{\it inter alia} lags and have compared our results to the {\it IUE}
fluxes extracted from the far-UV continuum, He\,II $\lambda$1640, and 
several resonance lines.

  A comparison of He\,I $\lambda$5875 and $\lambda$6678 line profiles 
during the peak of the infall activity suggests that differences in the 
development of the blue wing at this time are due to heating and a 
short-lived formation of an optically thin layer above the region 
compressed by the infall. This discovery and the well-known decreases in 
equivalent widths of the C\,II doublet at the two shock 
phases leads us to suggest that shock heating flattens the atmospheric 
temperature gradient, whether it is the infall shock preferentially
heating of the upper atmospheric layers from infall, or the pulsational
wave shock, which takes on an isothermal character as it emerges into
the more tenuous upper photosphere.

  Except for evidence of wind in the far blue wings of the UV resonance 
lines, we find no evidence for a shock delay arriving at different
regions of line formation of the photosphere (i.e., a ``Van Hoof effect').
Phase lags attributed by some former observers may be false indicators 
arising from varying degrees of desaturation of multiple lines, such as
for the red He\,I lines. In addition, an apparent 
lag in the equivalent width curve of lines arising from less excited atomic 
levels could instead be caused by post-shock cooling, followed by a 
rebound shock, as suggested by subtle variations in the photospheric 
$\lambda$1640 and UV continuum flux curves.



\end{abstract}


\section{Introduction}

 The $\beta$\,Cephei variable BW\,Vul (HR\,8007, HD\,199140; B1\,V to B2\,III) 
is in kinematic terms the largest amplitude pulsator known in the Galaxy. 
Its fundamental radial pulsation mode (Stamford \& Watson 1981, Aerts 
1995) is so strongly excited as to produce discontinuous ``standstill" 
features in the star's light curve and, immediately following this, 
a longer standstill in the radial velocity curve as well.
These features result from highly nonlinear processes associated with 
upward propagating pulsation waves. These waves emerge into the photosphere 
as highly supersonic shocks. 
During the pulsation cycle, the optical line profiles remain in absorption 
but undergo extreme variations in shape and velocity. 
Equivalent width variations are also noticeable at certain phases. 
In the often-used convention that $\phi$ = 0 occurs at light maximum (minimum 
radius), the radial velocity standstill becomes centered at $\phi$ = 0.98-1.00. 
Line profiles exhibit double lobes at phases just before (centered at $\phi$ 
$\approx$ 0.90), and during some cycles just after ($\phi$ $\approx$ 0.06)
the velocity standstill phase (e.g., Mathias et al. 1998). Adding to the
complexity of description, the
radial velocity curve is somewhat sensitive to the method of measurement, 
the spectral and temporal resolution of the observation, and especially
to the momentary pulsational amplitude of the star, for the amplitude of 
the pulsations fluctuate by several percent from night to night (Crowe 
\& Gillet 1989, Aerts et al. 1995, Mathias et al. 1998, Garnier et al.
2002). The equivalent 
widths of some metal lines vary with phases as a function of excitation 
potential (Furenlid et al. 1987). The finite signal-to-noise ratio 
and temporal sampling frequency of the {\it 
International Ultraviolet Explorer (IUE)} observations set practical 
limits on the otherwise considerable complementary UV information that 
they offer to optical spectra. 

  Despite these observational limitations, important effects of the star's 
pulsation cycle are readily visible on the atmosphere. One of these is
a variation of the effective surface gravity and especially the instantaneous 
``effective temperature" of the star during the cycle. Recently, 
Stickland \& Lloyd (2002) have compared flux variations at a range of 
wavelengths from the far-UV to the near-UV to show that the effective 
temperature varies from 20,000\,K to 25,000\,K during the cycle. 
Temperature variations this large may well cause observable modulations
in the mass loss and X-ray luminosity (cf. Cohen 2000) of the cycle.

   Historically, controversy has surrounded the interpretations of 
the profile and strength variations caused by shock waves moving through 
the atmosphere of BW\,Vul. One of these is the so-called ``Van Hoof effect," 
named after its primary discoverer (Van Hoof \& Struve 1953). This effect 
is the purported phase lag between the velocity curves extracted from lines 
formed at different atmospheric depths. This is thought to be the result of
the finite travel time required for a pulsational shock wave to move up
from one region of line formation to another. In the most recent such 
report, Mathias et al. (1998) reported that double line profiles 
of various lines observed near $\phi$ $\approx$ 0.9 and sometimes 0.1 
can exhibit equal blue-red strengths at slightly different phases.
A related issue is the cause of the line doubling itself. 
Odgers (1955) and Goldberg, Walker, \& Odgers (1976) first attributed the 
velocity discontinuities to  atmospheric absorptions just below and above 
the shock.
These authors argued that as an upward-propagating pulsation wave breaks 
into a shock it accelerates the line forming regions of the atmosphere 
from the lower photosphere, thereby creating a density discontinuity
with respect to the lower photosphere. The semi-ejected ``shell" coasts 
to some maximum height and returns ballistically to these lower strata.
In contrast, Young, Furenlid, \& Snowden (1981; YFS) suggested that 
turbulence and pressure effects are the chief causes of the profile 
doublings at these phases. In their picture the standstills are caused 
by line formation in both a lower stationary atmospheric region as well 
as an infalling zone rendered transparent by its lower density. 
All these studies have pointed to the extended displacements of the 
line formation region (several percent of a stellar radius) as well as 
its virtual free fall from maximum to minimum radius. 

  In the most recent kinematical description, Mathias et al. (1998) 
and Garnier et al. (2002) have summarized the present consensus 
that there are two shocks per cycle. The first, ``pulsation," shock 
is the result of the evolution of the upward-propagating 
wave which grows in amplitude from the envelope where it is excited. 
As it emerges into the atmosphere at $\phi=0.1$, 
this shock has a moderately high Mach (5-7) number, as referenced 
by the velocity ``discontinuity" just prior to the velocity 
standstill. A subsequent, ``infall," shock, 
occurring 0.8 cycles after the first,
is due to the extreme compression 
of the upper atmospheric strata as they fall back and catch up to the 
slower moving layers of the lower photosphere. In this picture the line
profiles exhibit double lobes during the main (and often infall) shock
because of the velocity jump associated with it. Mathias et al. also note
that because the density of the atmosphere decreases monotonically outwards, 
the infalling region cannot be described as a disconnected shell. In
addition, Mathias et al. suggested that shock progresses inward in terms 
of absolute (Eulerian; radius from star center) coordinates even as it 
moves outwards in mass. Thus, their
description reconciles the idea expressed by several previous authors that 
two outward-moving shocks per cycle
propagate up through the atmosphere. In the past the infall
shock, which forms at $\phi$ $\approx$ 0.90, has been mistaken 
for a reflection of a shock from the previous cycle off an interior
density gradient discontinuity. In this study we adopt the view 
of Mathias et al. that this shock is a natural consequence of infall, 
and that any earlier reflected shock is likely to be damped 
within the star, rendering it invisible at the surface.

  The elusiveness of even a qualitative interpretation of the shocks
in BW\,Vul has slowed the necessary development of self-consistent 
radiative hydrodynamical models. Early on,
Stamford \& Watson (1978) assumed that a large-amplitude velocity piston
at the base of the atmosphere developed into a thin, isothermal shock 
as it progressed through the line formation region. Using this 
dynamical model atmosphere, they constructed line profiles of
Si\,III $\lambda$4552 at several key phases in the cycle. Profiles
at phases we now call the infall shock exhibited line doubling (albeit
over only a brief interval).  In subsequent work Stamford \& 
Watson (1981) placed a large, adiabatic sinusoidal velocity 
variation at the base of a gray model atmosphere and demonstrated 
that an isothermal shock developed in the line formation region. 
Although they did not compute line profiles in these simulations, Stamford 
\& Watson stated that they anticipated that the shock would produce line 
doubling during the shock passage. The 1978 Stamford \& Watson paper 
to date represents the only attempt to compute the line transfer for a 
spectral line in a moving model atmosphere appropriate to BW\,Vul. Recently, 
Owocki \& Cranmer (2002) have developed hydrodynamic models that roughly
simulate the velocity and light variations of the pulsating star. 
Their models assume strong outgoing shocks in realistic atmospheres.
The shocks begin in the envelope as a large-amplitude pulsation wave, break 
into a small-amplitude shock in the lower photosphere, and evolve into
an isothermal, large-amplitude (60$\times$ in density) shock in the wind. 
Owocki \& Cranmer's results show that the shock has the 
effect of flattening the atmospheric density gradient well out into 
the wind, thereby explaining the large range in phase over which the shock
can be traced in the UV resonance lines. 
As these waves emerge into the wind, they can impact slower flows, causing 
reverse shocks and the formation of Discrete Absorption Components (DACs) 
in the far blue wings, indeed as observed in the Si\,IV and C\,IV lines 
(Burger et al. 1981, Blomme \& Hensberge 1985, Massa 1994). 

 The idea of a strong shock front moving through the atmosphere is of
course not limited to $\beta$\,Cephei variables. In fact, the concept 
of a vertically propagating ``hot front" was first used by Merrill (1955), 
referring to line doubling in both absorption and emission lines in long
period variable stars. While this still represents the common interpretation 
of this spectroscopic phenomenon in the LPV stars (e.g., Alvarez et al. 
2000), the idea of shocks heating the atmosphere have not taken root in 
most discussions of the dynamics of the BW\,Vul atmosphere. This is 
perhaps because emission components have not been observed in BW\,Vul's 
spectrum. In fact, the only evidence given for shock dissipation and 
heating appeared rather recently, in Furenlid et al.'s (1987) 
discussion of variable O\,II lines. The subject has been virtually 
ignored since, and we will revive this discussion in this paper.

  The present paper was motivated partially by our impression that an 
understanding of the shock wave properties has been hampered by 
uncertainties in spectroscopic measurements and interpretation.
For example, YFS and Crowe \& Gillet (1989) 
found widely different  {\it average} equivalent widths for the important C\,II
$\lambda\lambda$6578-82 lines. Yet the ``large" values found by YFS 
during the distension phases formed the basis of their conclusion that 
line doubling and strengthening is caused by changes in atmospheric 
continuous opacity. This suggestion though oft-quoted has gone untested. 
We also realized in planning our program that the existence 
of a Van Hoof effect could be tested by comparing the responses of the
red He\,I $\lambda$5875 (triplet) and $\lambda$6678 (singlet) lines, which 
probe slightly different column lengths through their $gf$ difference, 
and also volume densities because of the triplet line's mild sensitivity
to density through the partial metastability of its lower level. 
These lines have heretofore not been monitored in BW\,Vul.
Other significant but so far unremarked lines in the red region 
are the Si\,II $\lambda$6371 and Si\,III $\lambda$5737 features, which
together provide a measure of variations in the atmosphere's ionization
equilibrium. The echelle data we have obtained sample formation conditions 
of many lines at the same time and thus can provide accurate phasing 
information from this variable-amplitude pulsator. This simultaneity
permits us to remove past ambiguities in correlating behaviors of various
lines at different epochs. A second reason for undertaking this study is 
the availability of largely unanalyzed {\it IUE} data from several cycles
at nearly the same epoch. This dataset permits us to tie together the 
resonance lines and high-excitation optical lines formed deeper in the 
atmosphere at a large number of phase reference points in the pulsation
cycle.

\begin{table}
\caption{Journal of Year 2000 Observations (HJD - 2400000)}  
\label{tab_obsdates}
\begin{center}
\begin{tabular} {c|c|c} \hline
Night 1 (Sept. 17/18) & Night 2 (Sept 18/19) & Night 3 (Sept. 19/20) \\ \hline
    51805.574  	 &   51806.560         &    51807.568  \\
    51805.579    &   51806.564         &    51807.578  \\ 
    51805.583  	 &   51806.569   &	    51807.585  \\ 
    51805.588  	 &   51806.574   &	    51807.593  \\ 
    51805.593  	 &   51806.579   &	    51807.599  \\ 
    51805.597  	 &   51806.584   &	    51807.604  \\
    51805.602  	 &   51806.589   &	    51807.609  \\
    51805.607  	 &   51806.593   &	    51807.613  \\
    51805.612  	 &   51806.600   &	    51807.617  \\
    51805.616  	 &   51806.606   &	    51807.621  \\
    51805.621  	 &   51806.612   &	    51807.651  \\
    51805.625  	 &   51806.620   &	    51807.655  \\
    51805.629  	 &   51806.626   &	    51807.659  \\
    51805.633  	 &   51806.632   &	    51807.754  \\
    51805.637  	 &   51806.638   &	    51807.759  \\
    51805.643  	 &   51806.646   &	    51807.764  \\
    51805.649  	 &   51806.653   &	    51807.767  \\
    51805.653  	 &   51806.657   &	    51807.771  \\
    51805.657  	 &   51806.662   &	    51807.775  \\
    51805.661  	 &   51806.666   &	    51807.781  \\
    51805.665  	 &   51806.671   &	    51807.785  \\
    51805.672  	 &   51806.677   &	    51807.789  \\
   51805.676  	 &   51806.681   &	    51807.793  \\
   51805.680  	 &   51806.686   &	    51807.797  \\
   51805.684  	 &   51806.691   &	    51807.801  \\
   51805.688  	 &   51806.696   &	    51807.805  \\
   51805.692  	 &   51806.701   &	    51807.809  \\
   51805.696  	 &   51806.705   &	    51807.813  \\
   51805.700  	 &   51806.710   &	    51807.817  \\
   51805.704  	 &   51806.715   &	    51807.821  \\
   51805.708  	 &   51806.720   &	    51807.826  \\
   51805.715  	 &   51806.724   &	    51807.830  \\
   51805.719  	 &   51806.728   &	    51807.834  \\
   51805.723  	 &   51806.732   &	    51807.838  \\
   51805.727  	 &   51806.736   &	    51807.843  \\
   51805.731  	 &   51806.740   &	    51807.849  \\
   51805.735  	 &   51806.747   &	    51807.855  \\
   51805.739  	 &   51806.751   &	    51807.860  \\
   51805.743  	 &   51806.755   &	    51807.865  \\
   51805.747  	 &   51806.759   &	    51807.871  \\
   51805.751  	 &   51806.764   &	    51807.876  \\
   51805.759  	 &   51806.770   &	    51807.882  \\
   51805.763  	 &   51806.774   &	    51807.887  \\
   51805.768  	 &   51806.778   &	    51807.893  \\
   51805.772  	 &   51806.782   &	    51807.898  \\
   51805.776  	 &   51806.787   &	    51807.904  \\
   51805.781  	 &   51806.791   &	    51807.909  \\
   51805.785  	 &   51806.795   &	    51807.916  \\
   51805.789  &	    51806.798  	  &      -     \\
   51805.793  &    51806.803  	  &      -     \\
   51805.800  &    51806.808  	  &      -     \\
   51805.810  &    51806.813  	  &      -      \\
   51805.815  &    51806.818  	  &      -      \\
   51805.819  &    51806.822  	  &      -       \\
   51805.823  &    51806.827  	  &      -     \\
   51805.832  &    51806.832  	  &      -      \\
   51805.837  &    51806.837  	  &      -       \\
   51805.841  &    51806.842  	  &      -       \\
   51805.846  &    51806.847  	  &      -        \\
   51805.851  &    51806.851  	  &      -        \\
     -         &   51806.856  	  &      -        \\
\hline
\end{tabular}
\end{center}
\end{table}

\begin{table}
\caption{Summary of Atomic Data for Lines in McDonald Spectra } 
\label{tab_atdata}
\begin{center}
\begin{tabular}{cccc} \hline
Wavelength & Ion & $\chi$ (eV) & log\,$gf$ \\
\hline 
5606.09  & S\,II  & 13.8 & 0.16 \\ 
5639.980 & S\,II  & 14.1 & 0.33 \\
5640.314 & S\,II  & 13.8 & 0.15 \\
5646.979 & S\,II  & 14.1 & 0.11 \\
5648.070 & C\,II  & 20.8 & -0.45 \\
5639.980 & S\,II  & 14.1 &  0.33 \\
5640.314 & S\,II  & 13.8 &  0.15 \\
5659.956 & S\,II  & 13.7 & -0.07 \\
5662.460 & C\,II  & 20.8 & -0.27 \\
5666.629 & N\,II  & 18.5 &  0.01 \\
5676.017 & N\,II  & 18.5 & -0.34 \\
5679.558 & N\,II  & 18.6 &  0.28 \\
5686.213 & N\,II  & 18.6 & -0.47 \\
5696.603 & Al\,III & 15.7 & 0.23 \\ 
5710.766 & N\,II  &   18.6 & -0.47 \\
5722.730 & Al\,III &  15.7 & -0.07 \\
5739.734 & Si\,III & 19.8 & -0.160 \\
 5747.300 & N\,II  & 18.6 & -1.020 \\
 5833.938 & Fe\,III & 18.6 & 0.616 \\
 5875.615 & He\,I  &  21.0  & 0.41 \\
 6247.178 & Al\,II &  16.6 & -0.20 \\
 6346.859 & N\,II  &  23.3 & -0.86 \\
6371.371 & Si\,II &  8.2 & 0.00 \\
6379.617 & N\,II  &  18.5 & -0.92 \\
 6562.801  &  H\,I  &  10.20 & -0.69 \\
6578.052 & C\,II  & 14.5 & 0.12 \\
6582.882 & C\,II  & 14.5 & -0.18 \\
6678.154 & He\,I  & 21.3 & 0.33 \\
\hline
\end{tabular}
\end{center}
\end{table}

\section{Observations } 
\label{obs}

\subsection{Observations and reduction details }

  The optical data for this study were obtained on the nights of 2000
September 19--21 with the Sandiford echelle spectrograph (McCarthy et 
al. 1993) attached by fiber optics to the cassegrain focus of the 
Struve 2.1-m telescope at McDonald Observatory. The cross-dispersing 
prism in this instrument was rotated to select a central wavelength of
6120\AA\ and to include a nearly continuous wavelength range (22 orders) 
of $\lambda\lambda$5510-6735 on a 1200 $\times$ 400 CCD detector. 
This configuration resulted in a spectral resolution of 45\,000 and a 
pixel spacing of 2.8 km~s$^{-1}$ pix$^{-1}$. 
Signal-to-noise ratios of 200--300 were typically attained
in integration times of 5--6 minutes, except at high airmass when they
were increased to 7--8 minutes. (The profiles displayed in 
Fig.~\ref{fig_heprof} of this paper were exposed 5--6 minutes.) 
Flat and comparison lamp spectra taken at various times of the night showed 
that the continuum and wavelength stability of the spectrum was robust. 
A table of the mid-observation times for the three nights is compiled 
in Table\,1, while a line list of identified features in our spectra is 
given in Table\,2. The wavelengths and excitation potentials listed are 
obtained from the Kurucz line library (Kurucz 1990).

  The spectra were kindly reduced (background-subtracted, extracted, and 
flatfielded) by Dr. Chris Johns-Krull using computer codes described by 
Hinkle et al. (2000) and Piskunov \& Valenti (2002). Wavelengths were 
determined by an interactive graphics package that allowed a solution
simultaneously in the echelle and cross-dispersion axes (Valenti, priv. 
comm.).  Solutions were obtained by minimizing residuals between laboratory 
and observed Th and Ar line wavelengths and iteratively rejecting outliers. 
Corrections for terrestrial orbital and rotational velocities were made.
Rectification of echelle orders was performed by an interactive polynomial 
fitting procedure. Orders were then spliced at wavelengths for which flux
contributions of adjoining orders were equal.
We measured the radial velocities of strong lines by cross-correlating the
lines against a reference line profile observed near $\phi$=0.2. Profiles
at this phase are approximately symmetric and exhibit an approximately
mean width and thus lend themselves to comparison with profiles of extrema 
phases. We determined true equivalent widths of these lines and other 
analyzed features in the McDonald spectra by an interactive computer 
algorithm tailored for this application. The program uses input wavelength 
ranges over which both the continuum and the line's absorption profile 
are to be extracted. The continuum level is then fixed by a specified 
``Nth percentile brightest flux" among candidate fluxes in the continuum 
window. The value of N, typically $\approx$80\%, can depend on the 
presence of nonstellar features but is well suited to modification 
interactively when such unwanted features appear as telluric lines, 
cosmic rays, or fringing. 

\subsection{IUE Spectra}
\label{iues}

   In order to avoid uncertainties in the absolute calibration of fluxes, 
we made use of 38 high-dispersion echellograms obtained 
with the {\it SWP} camera through the large aperture during monitoring 
campaigns on the star in October and November 1994. ``NEWSIPS" extractions 
were downloaded from the MAST\footnote{ Multi-Mission Archive at Space 
Telescope Science Institute, under contract to NASA.} web-based archives.  
Absorption {\it line strength indices} (LSI) were then calculated by ratioing the 
total {\it net} (uncorrected for ripple distortion) flux in a narrow band 
centered at line center to the total net flux in the parent echelle order. 
Such indices 
are not true equivalent widths, but they are directly proportional to
them and increase with absorption strength. Because these indices are
insensitive to continuum placement, and (as defined herein) independent 
of errors in blaze function (``ripple'') correction, they are unambiguous 
measures of the absorption for prescribed velocity limits, and they are 
particularly accurate differential measures of line strength differences 
for an {\it IUE} time series. The line fluxes were measured between 
velocity limits of generally $\pm{1.0}$\,\AA\, of line center, where 
the radial velocity of the star was referenced at --10 km~s$^{-1}$. 
Because the radial velocity of the star changes dramatically through the 
cycle, we first determined spectrum-to-spectrum shifts in pixel space 
before co-adding spectra in a co-moving frame.

\subsection{Radial Velocity Ephemeris}

  Although there is unanimity that BW\,Vul is monoperiodic (with 
P $\approx$ 0.201043), some controversy has surrounded the ``drifting"
of its pulsation period derived from datasets of different epochs. 
Various authors have suggested ephemeris corrections for a
quasi-evolutionary lengthening, light travel time across a binary, 
and both random and discontinuous changes for unspecified reasons.
In past years support has built for the binarity solution with an 
orbital period near 34 years (Pigulski et al. 1993, Odell 1994, Horvath,
Gherega \& Farka 1998), although it appears possible that small random 
changes can also alter the phase zeropoints from time to time (Sterken 
1993). We have adopted the binary + secular-lengthening solution of 
Horvath et al. for the light minimum phases. Light minimum is the recommended 
new benchmark because it can be determined with greater precision (Sterken 
1993). To these calculated phases, we have added 0.48 cycles (Furenlid et 
al. 1987, Stickland \& Lloyd 2002) to reference them to the traditional 
zero at optical light-maximum. This is also the
approximate midpoint of the radial velocity standstill. This ephemeris
agrees to within 0.015 ($\pm{0.02}$) cycles of the midpoint occurrence
of the standstill in our McDonald data. The relative phases of the 1995
{\it IUE} epoch did not match as well, giving a difference of 0.09 cycles
from this ephemeris relative to the standstill occurrence found by 
Stickland \& Lloyd. We have referred our $\phi$ = 0 fiducial to the 
Stickland \& Lloyd phase $\phi$ = 0.10, which takes into account that the 
latter authors tied their zeropoint to radial velocity maximum instead 
of the usual velocity standstill criterion.\footnote{Note that the 
photometric standstill occurs just before the onset of velocity standstill 
and is much shorter. These respective standstills last about 0.03 and 
0.15 cycles (see e.g., Furenlid et al. 1987).}

\section{Results}
\label{res}

\begin{figure}
\includegraphics[height=84mm,angle=90]{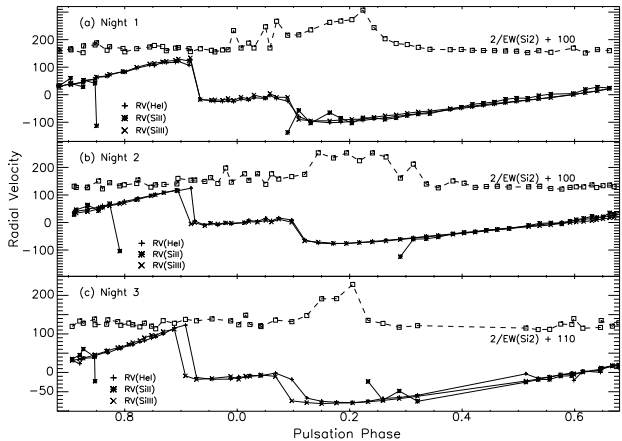}
\caption{Radial velocity in km\,s$^{-1}$ determined from cross-correlations
on the He\,I $\lambda$6678 (pluses), Si\,III $\lambda$5740 (crosses)
and Si\,II $\lambda$6371 (asterisks)
lines of BW\,Vul for all three nights of this study, 2000 September 19-21.
Note that the He\,I
velocities have been rescaled by factors of 1.10, 1.05, 1.05 to match the
Si\,II line velocity amplitudes. The squares at the top of each panel
denote the reciprocal (with offsets of $\sim$+100) Si\,II equivalent widths;
these could not be measured for phases near zero when the two shocks occur.
}
\label{fig_rvb}
\end{figure}

\subsection{Radial velocities}
\label{radv}

  A large number of radial velocity curves have been discussed for 
BW\,Vul from as many optical datasets. We decided to focus our work
on the radial velocities of the red helium lines and primarily for the 
purposes of checking the phase zeropoint and to search for differences
in amplitude among the cycles on the three nights of our observations.
We determined an instrumental systemic velocity of the star by 
comparing the centroid wavelength of symmetric line profiles
with nearby Th-Ar features in our emission spectra and correcting
them for the Earth's orbital and rotational velocities. We then averaged
all the velocities around their respective cycles. The resulting mean
heliocentric radial velocities found for BW\,Vul from the three nights were 
--7.7, --10.3, and --11.4 km\,s$^{-1}$. The resulting mean of --9.7 km\,s$^{-1}$ 
is in excellent agreement the mean value of --9.2 km\,s$^{-1}$ given by 
Mathias et al. In Figure~\ref{fig_rvb} the radial velocities are shown for 
the He\,I $\lambda$5875 for all the three nights. It was possible to 
measure the equivalent widths of the lines arising from highly excited 
atomic levels, but these often were both weakened and broadened to 
invisibility during the critical shock passage phases. 
Radial velocities are also given in Fig.~\ref{fig_rvb}
for Si\,III $\lambda$5740 and Si\,II $\lambda$6371
(the latter can reliably be measured only in the phase ranges 0.1 $<$ $\phi$ 
$<$ 0.85). The silicon lines have velocity amplitudes 5-10\% 
larger than the helium line curves and even more striking discontinuities
at the beginning at end of standstill. The H$\alpha$ velocity
curve (not shown) has even a slightly smaller amplitude and less steep
``discontinuities" than the He\,I line curves do. 
We believe that these are effects of the far wings of the helium and 
hydrogen lines, which tend to broaden the cross-correlation function 
and produce artificially small velocity differences for the line core.
Aside from this anomaly, the silicon line velocities are similar to the
helium ones. All these transitions have moderate to high excitation potentials 
($>$20 eV) and are saturated. 
Thus, in static atmospheres without the imposition of shocks 
one expects them to be formed in the lower photosphere close to the 
continuum formation region. Thus it should not be surprising that they do 
not show phase lags (Van Hoof effects). By comparison, we also exhibit in 
the upper region of the three panels of Figure~\ref{fig_rvb} the equivalent
widths of the Si\,II $\lambda$6371 line over those phases outside the 
shock intervals - these are the times when the line was strong enough 
that reliable centroid positions could be measured. 
On Nights 2 and 3 one sees that the phases of velocity maximum extend 
slightly longer in the Si\,II feature than in the He\,I line, causing
a slight delay in the onset of the standstill for the He\,I line curves. 
We believe that 
this is an artifact of a relatively ``late" desaturation of the blue lobe 
of the weaker Si\,II line, causing the line's centroid velocities to 
remain at negative velocities for a longer time and to later phase 
than the other lines measured (see $\S$\ref{nub}). 
We note also that the average radial 
velocity of the two red He\,I lines during standstills of each of
the three nights was --8.3, +2.8, and --11.6 km~s$^{-1}$, respectively,
giving a mean of --6 km~s$^{-1}$. This is very nearly the systemic velocity
of the star, a result similar to that found by Mathias et al. (1998).
Finally, differences among the two He\,I lines and the Si\,II line averaged
$\pm{4}$ km~s$^{-1}$, so it is likely that there are minor cycle-to-cycle 
differences in the mean velocity of the standstill.

   Figure~\ref{fig_rvb} provides a useful means of correlating the He\,I line
velocities with equivalent width changes for this and other lines
during the cycle. Notice first 
the clearly defined standstill of the velocity curve, lasting some 
0.16 cycles. This feature can be used to check the phase zeropoint given in
past ephemerides. The features of the standstill are found to be similar in 
form, whether determined from a line-centroiding technique or from fitting 
two gaussians when they appear. Additionally, the feature also shows 
a shallow positive slope, as several other investigators have noted who
used a different (double-gaussian) velocity measurement criterion. 
The standstill velocities are near zero or slightly negative. 
In addition, Fig.\ref{fig_rvb} 
indicates that the positive velocity amplitudes, and phases of the 
quasi-discontinuity leading to the standstill, can change from one 
cycle to the next. We can speculate that these differences reflect the
variable pulsation shock strengths from the preceding cycles.

\begin{figure}
\includegraphics[height=84mm,angle=90]{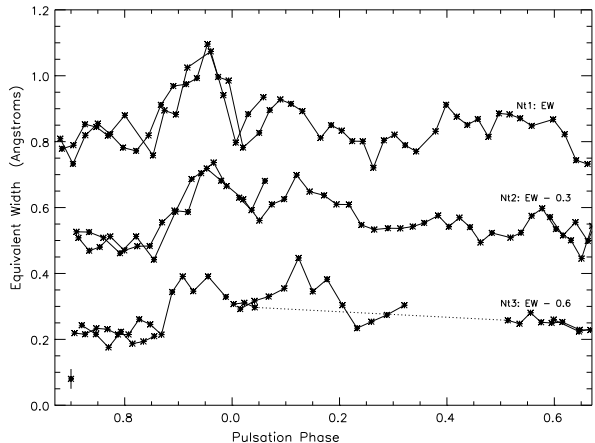}
\caption{ Equivalent widths of He\,I $\lambda$6678 with phase 
for the three nights of study; offsets of -0.3\AA\ and -0.6\AA\ are
introduced for visual clarity. Approximate error bars are indicated.
}
\label{fig_heew}
\end{figure}

\subsection{Behavior of the Excited Optical Lines}
\subsubsection{The red He\,I lines}
\label{redhe}

   The accurate measurement of equivalent widths of lines in the BW\,Vul 
spectrum is challenging because the positions, widths, and core depths 
vary dramatically over the cycle. We began our study by investigating 
equivalent widths for the $\lambda$5875 and $\lambda$6678 lines. 
These He\,I transitions are, respectively, analog triplet and 
singlet 2P-3D transitions, and each has excitation potentials of 21 eV. 
Although variations of the neutral helium line have not yet been studied 
in this star, their importance cannot be overstated because of the
lines' sensitivity to atmospheric heating. Furenlid et al. (1987) reported 
unambiguous evidence for a temperature rise from increases in the ratio 
of a pair of O\,II and Fe\,III lines during the velocity-standstill phase.
Figure~\ref{fig_heew} shows the variation of the true equivalent 
widths for the He\,I $\lambda$5875 and He\,I $\lambda$6678 lines for 
all three nights. These plots exhibit generally two maxima, the stronger
of the two centered at the occurrence of the infall shock at $\phi$$\approx$
0.90-0.95. The equivalent width ratio of these lines is 1.13$\pm{0.03}$ 
outside the ``windows" of the two shock intervals. Because the ratio of
their atomic $gf$'s is two, the observed ratio indicates that the features 
are very optically thick. The measured ratio is almost the same during the 
passage of the pulsation shock, but it increases to a mean value of 
1.23$\pm{0.08}$ during the infall shock. 

\begin{figure}
\includegraphics[height=84mm,angle=90]{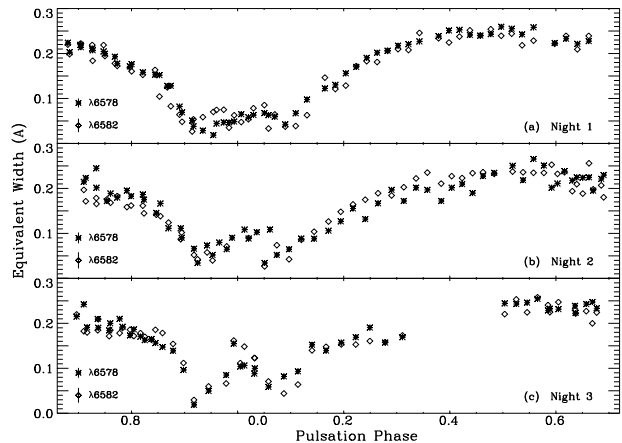}
\caption{ C II $\lambda$6578, $\lambda$6583 equivalent width curves for
the three nights in this study. The values for $\lambda$6583 are scaled 
by factors of 1.3, 1.1, and 1.2 for the three nights, respectively,
to match the $\lambda$6578 data.
}
\label{fig_c2ew}
\end{figure}

\subsubsection{The C\,II $\lambda$6578,  $\lambda$6583 doublet and 
high-excitation lines}

  The C\,II $\lambda$6578, $\lambda$6583 doublet is located close to the 
H$\alpha$ line and arises from a similarly excited level of 14.4 eV. It 
has thus long served as a conveniently accessible temperature probe for
atmospheres of variable B stars. We measured the true equivalent 
widths of each of these lines in the same manner as the He\,I lines. 
These are exhibited in Figure~\ref{fig_c2ew}, where we have represented 
the two lines by different symbols and rescaled the equivalent widths of
$\lambda$6583 to the slightly larger ones of $\lambda$6578. Some of the
scatter at like phases arises from measurements at adjacent pulsation 
cycles. The $gf$ ratio of the components is 2. The observed ratios 
do not show evidence of variations during the cycle, and their nightly 
means, 1.28, 1.14, and 1.16, show clearly that even though the lines are 
comparatively weak they are quite optically thick. 

  Our C\,II doublet curves in Fig.~\ref{fig_c2ew} exhibit well defined minima 
that coincide in phase with the broad minimum found by Crowe and Gillet 
(1989), except that the latter authors' data do not hint at a separation
in phase between the two individual shocks.
The C\,II curves are also very similar to those shown for the residual 
intensity of the H$\alpha$ core discussed by Crowe \& Gillet (1989). 
We have confirmed this behavior for the core of this line in our data 
and have also found that the equivalent widths extracted from small 
wavelength windows centered on the core exhibit a similar behavior.
The effect rapidly disappears and even inverts as one includes more 
and more contribution of the H$\alpha$ wings in the window.
These results confirm that the variations are produced by localized 
atmospheric strata. Other
moderate to high excitation lines such as N\,II $\lambda$5710 ($\chi$ = 
18.5 eV) and S\,II $\lambda$5640 (13.8 eV) produce variations similar to 
the C\,II doublet, but the intrashock maxima for them are not always so 
clearly separated in phase. For phases outside the two shock intervals, 
many of the lines in our sample broaden and fade to below our detection 
threshold.

\subsubsection{Si\,II $\lambda$6371 and Si\,III $\lambda$5740 lines}
\label{hilo}

  The Si\,II $\lambda$6371 and Si\,III $\lambda$5740 lines are important
because they arise from atomic levels having the largest combined excitation 
and ionization potentials of all the lines in our optical coverage, save
the He\,I lines. 
In addition, their combined responses furnish information on changes in 
the ionization equilibrium in this atmosphere during the pulsation cycle.  
In investigating the strengths of these lines, we found first that the 
Si\,III $\lambda$5740 line shows only mild increases of $\sim$10--50\%
in equivalent width from night to night during shock passage. 
This is because the ionization equilibrium of silicon is roughly 
balanced between Si$^{+1}$ and Si$^{+2}$. However, as the dashed lines at
the top of each of the panels 
in Figure~\ref{fig_rvb} show (plotted are the reciprocals of the line strength), 
the response of the Si\,II $\lambda$6371 line can be quite pronounced. 
During the passage of the second (pulsation) shock, the $\lambda$6371 broadens
and weakens so much that its velocity centroid cannot be reliably measured.
As indicated in Fig.~\ref{fig_rvb}, the phases of maximum weakening do not
coincide with the shock passage but rather are delayed by 0.10 cycles 
after the end of the standstill phase, when the strengths of the excited
lines are slowly decreasing. 

    Figure~\ref{fig_c2ew} gives an indication of the variations in shock 
heating and of the inequality of heating amplitudes between the two shocks.
One can see this, first, in the amplitude variations in the shocks (as 
judged by the sharpnesses and depths of the C\,II minima) and, second, in 
the variation of the strengths of the Si\,II line and other moderate 
excitation lines, e.g., S\,II $\lambda$5640. These diagnostics generally 
imply that the temperature increase associated with the pulsational wave 
is the stronger of the two shocks. This is contrary to the inequality of
the shock jumps (Fig.~\ref{fig_rvb}; see also Mathias et al. 1998). 
We also note for discussion below that 
the phase of equivalent width minimum is delayed by 0.07--0.08, both with 
respect to the line's radial velocity minimum (Fig.~\ref{fig_rvb}) and with 
reference to the C\,II equivalent width minimum at $\phi$ = 0.1.

\subsection{Ultraviolet lines}
\label{uv}

  Except for the He\,II 1640A line (``helium H$\alpha$," $\chi_{exc}$
= 41 eV), the UV lines used in this analysis are all resonance lines. 
To varying degrees the latter lines have components formed both in the 
static line forming region (upper photosphere) and the accelerating wind. 
We will treat these lines in order of likely formation depth, starting 
from the He\,II feature, and work our way up through ionization potential out 
into the wind.

\begin{figure}
\includegraphics[height=84mm,angle=90]{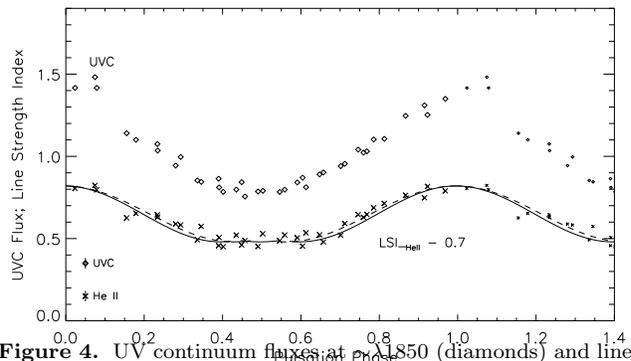}
\caption{ UV continuum fluxes at $\sim$$\lambda$1850 (diamonds) and
line strength index for He\,II $\lambda$1640 (crosses). The latter,
denoted LSI, is defined in $\S$\ref{iues}.
UVC fluxes are plotted relative to the mean, but 0.7 units are
subtracted from the He\,II LSI to separate the curves. Symbols for 
data above $\phi$ = 1.0 are repeated and rendered small. The dashed 
line through the He\,II curve is a reference sinusoid:
solid line is a compressed/stretched curve around the sinusoid (see
text). Note the
slight bump in the He\,II curve for data in the $\phi$ range 0.2--0.28 
and the weak dip at 0.15--0.2. Approximate error bars are shown.
}
\label{fig_he2}
\end{figure}

\subsubsection{He\,II $\lambda$1640 }

  Figure~\ref{fig_he2} depicts the extracted UV continuum (UVC; representing
$\lambda\lambda$1800-1905) light curves from the 37 available 
large-aperture {\it IUE/SWP} spectra obtained in 1994, together with the 
extracted line strength index created from ratios of fluxes within about 
$\pm{1.6}$\,\AA~ of He\,II line center to all the other net (ripple 
uncorrected) fluxes in the echelle order containing the line. The detailed 
shape of the UVC curve is in excellent agreement with the two plots 
constructed by Stickland \& Lloyd (2002) from essentially the same data. 
The curve shows a broad, asymmetric maximum which may be an underresolved 
rendition of two peaks apparent in the line strength curves at $\phi$ = 
0.9 and 0.1. This double-peaked structure is completely unresolved 
in optical light curves (cf. Fig.\,3 of Stickland \& Lloyd 2002). 
Note also that the scatter is small and does not seem to reflect the 
obvious cycle-to-cycle differences at some phases in the radial velocity 
curves. The He\,II $\lambda$1640 
absorption curve is morphologically very similar to the UVC curve. The
curves extracted from the blue and red halves of the profile are in turn
identical to one another. The similarity between the UVC and He\,II
curves is the result of the proximity of the depths of formation of
this line and the continuum.  However, the slight tendency of the He\,II 
curve to flatten at the maximum- and minimum-flux phases is likely caused 
by both the core and line (formed over different depth ranges) being 
sampled in our extraction. A slight bump in the He\,II curve at 
$\phi$ $\approx$ 0.20--0.28 is visible in Figure~\ref{fig_he2} and is also 
reproduced in the UVC curve. To clarify this bump and the weak dip occurring
at a phase just before it ($\phi$ = 0.15--0.20), we overplot two curves in 
Fig.~\ref{fig_he2}. The first is a simple sine curve (dashed line), 
and the second (solid line) is a sine curve that has been
compressed/stretched  in the first and second quadrants.
No matter how one chooses a fitted curve 
to the observed values (from three different pulsation cycles) 
the dip and bump features fall outside the error limits.
These weak features are also visible in the two UVC curves of Stickland 
\& Lloyd, based on 21 additional spectra and from data processed and 
extracted by different algorithms from ours. These are undoubtedly 
real, and we comment further on them in $\S$\ref{lag}.

\begin{figure}
\includegraphics[height=84mm,angle=90]{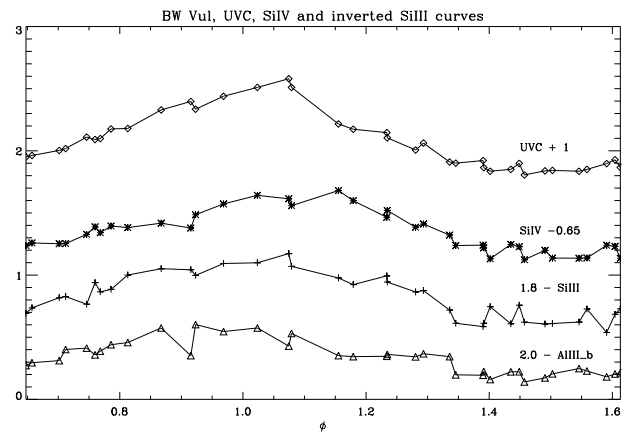}
\caption{ UV continuum fluxes at $\sim$$\lambda$1850 (diamonds) and
line strength indices for the 
Al\,III $\lambda$1855, Si\,III $\lambda$1206, and Si\,IV  $\lambda$1394
resonance lines. The line strengths are obtained from the central
profile within about $\pm{0.8}$ \AA\ of line center and refer to the
strength of the continuum. In the case of Si\,III and Al\,III the indices 
is inverted because the line responds oppositely from the other features.
Error bars are indicated.
} 
\label{fig_alsi}
\end{figure}

\begin{figure}
\includegraphics[height=84mm,angle=90]{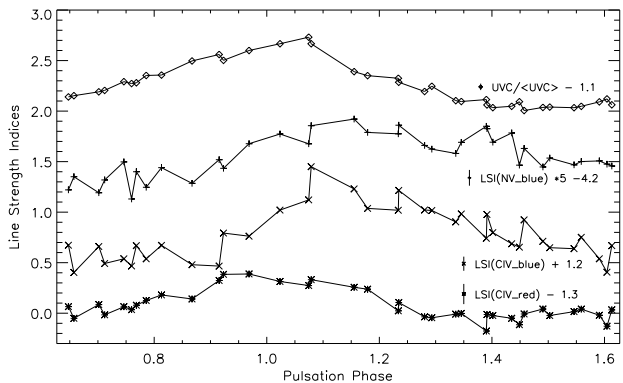}
\caption{ UVC fluxes (diamonds) and line strength indices for
Si\,IV $\lambda$1394, C\,IV $\lambda$1548, and N\,V $\lambda$1238,
either blue or red portions of the profile, as indicated. The windows
sampled are about $\pm{0.8}$\AA\ around line center. Error bars are 
indicated.
}
\label{fig_c4n5}
\end{figure}

\subsubsection{Resonance lines of moderately excited ions: Al\,III,
Si\,III, and Si\,IV}

  Figure~\ref{fig_alsi} depicts the UVC and absorption curves for the
lines Al\,III $\lambda$1855, Si\,III $\lambda$1206 and Si\,IV
$\lambda$1394. We have inverted the Al\,III and Si\,III curve in
these plots to facilitate a comparison among all four curves.
Curves extracted from the blue halves of the Si\,III and Si\,IV 
lines have 10--20\% larger amplitudes than the red halves but 
otherwise are similar. Note first that the Si\,III curve 
weakens (rises) more quickly than Si\,IV strengths during the phase 
interval 0.7--0.9. This is likely to be a consequence of Si$^{+3}$ 
being the dominant ion in the atmosphere, and Si\,III responding 
(decreasing in strength) more readily to heating that accompanies
the infall shock at $\phi$ $\approx$ 0.9. Secondly, note that as one 
proceeds from the Al\,III through the Si\,IV curves, the maximum 
absorptions of the excited ions extend over longer and longer intervals.
Figure~\ref{fig_c4n5} demonstrates that their long-lived character continues
for the C\,IV $\lambda$1548 and N\,V $\lambda$1238 (blue half only) 
features. For these lines one sees that this plateau extends to $\phi$ 
$\sim$ 0.4, which is some 0.3 cycles after the shock passes through the 
photospheric line formation regions of C\,IV and N\,V. Experimentation 
with extractions along the C\,IV profile indicates that as one includes 
more and more blue wing (wind absorption) in the measurement, the onset and 
extension of the maximum-strength phase shifts to later and later phases. 
Thus, this appearance of the shock as a long-lived feature in the wind 
is responsible for the positive phase shift of the ``C\,IV\_blue" and 
``N\,IV\_blue" curves relative to the ``C\,IV\_red" plot in this figure.

\section{Discussion} 
\label{disc}

\subsection{``Phase lags" reinterpreted }
\label{lag}

  The combination of optical and UV resonance line results requires a
picture which integrates the effects of the pulsation and infall-generated 
shocks. Figures~\ref{fig_he2}, \ref{fig_alsi} and \ref{fig_c4n5} record the differences 
in UV line responses with increasing ionization potential. Massa (1994) 
has depicted the acceleration of C\,IV, Si\,IV, and Si\,III in grayscale
{\it IUE} 1979 spectra. These spectra show that the wind features start 
0.1 cycles after the passage of the pulsation wave through the photosphere. 
The photospheric and wind components of C\,IV are at times co-mingled,
but otherwise there is no hint of a graduated response through the
photosphere. Rather, there seem to be separate responses from
two discrete  ``photospheric" and ``wind" regions.
In particular, continuum fluxes and lines of high excitation, 
such as He\,I, He\,II, and Si\,III, show no perceptible phase 
differences in their velocity or line strength responses. Even the 
line strength minima of the moderate excitation C\,II doublet 
($\chi$ = 14 eV) seem to coincide with the appearance of the shocks
just prior to and following the end of the velocity standstill. 
These lines collectively span a large range in excitation. 
Lines arising from less excited levels than 14 eV,
such as the resonance lines of Al\,III or moderate excitation
lines of Si\,III (Stickland \& Lloyd 2002), cannot be measured precisely
enough in {\it IUE} spectra to search for phase lags with respect to
the optical lines. In fact, we find that in order to see any obvious
indication of a phase lag after the passage of the excited lines one
must search in the far blue wings of the resonance lines.
Perhaps this should not be a surprise.
According to the Owocki \& Cranmer (2002) simulations, the UV absorption
features form over length scales much longer than the effective depth
of the static line formation region. Thus, there is ample column length
over which shock-induced features can develop.

   If there are no phase delays among the excited optical lines, what 
is one to make in Figure~\ref{fig_rvb} of the apparent
phase delay of 0.07--0.08 after the Si\,II $\lambda$6371 curve?
We believe the key here is that Si$^{+1}$ is a trace atmospheric ion,
which is therefore sensitive to temperature and {\it not} that its mean 
line formation region is so high in the atmosphere that it takes the 
shock a finite time to reach it. Consider as a more plausible circumstance
that the formation regions of most optical lines largely overlap. 
The so-called ``delayed" weakenings of the Si\,II feature may more 
easily represent the subsequent passage of a cooler medium. We have in 
mind the post-shock region which is cooler and more tenuous than the 
shock interface (Liberman \& Velikovich 1986). A strong post-shock 
rarefaction is clearly visible in the hydrodynamic simulations of Owocki
\& Cranmer (2002) and indicate a rough ``half-wavelength" of $\sim$0.25 
cycles. From this result one might anticipate the effects of a rebound 
shock having approximately this delay. Altogether, we can speculate that 
the marginally visible bumps in the UVC and He\,II strength curves at 
$\phi$ $\approx$ 0.23 are produced by a small temperature enhancement 
due to the rebound. In this picture the abrupt-appearing end of the UVC 
maximum would be the result of particles entering the cooled post-shock 
flow.

\begin{figure}
\includegraphics[height=84mm,angle=90]{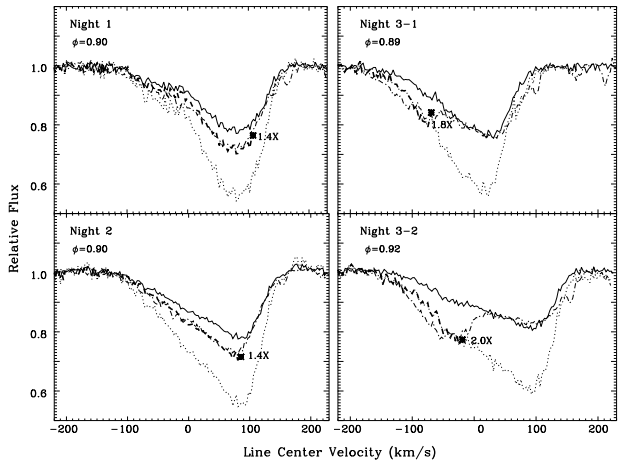}
\caption{Profile for the red $\lambda$5876 (dashed) 
and $\lambda$6678 (solid) He\,I lines, which have an atomic $gf$ 
ratio of 2:1, at a phase near $\phi$ = 0.90 on each of three nights. 
Also depicted are the $\lambda$6678 feature scaled by a factor of 2
(dotted line), as well as this same feature (thick dot-dashed) scaled by 
the indicated scale factor given next to the solid dot. This figure
suggests that the ``excess 
absorption" of the blue segment of both lines is due to optical thinness 
of the line at negative velocities. The lines seem to be strictly optically 
thin (ratio of 2) up to about --30 km~s$^{-1}$ on ``Night 3" (September 21). 
}
\label{fig_heprof}
\end{figure}

\begin{figure}
\includegraphics[height=84mm,angle=90]{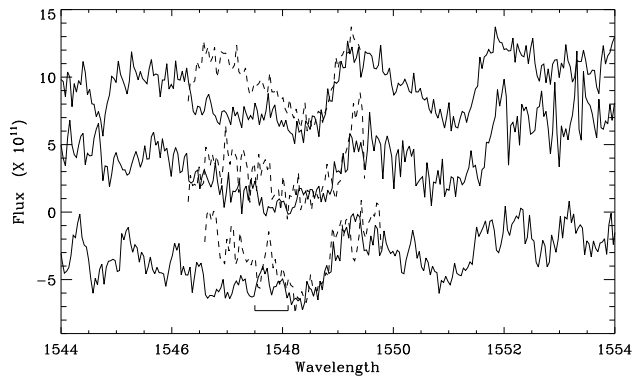}
\caption{ Montage of C\,IV profiles ($\lambda$1550 overplotted onto 
$\lambda$1548) at $\phi$ $\approx$ 0.9, showing the optical thinness of 
the {\it low velocity} blue wing at this phase. The observations, 
SWP\,52644-5 \& SWP\,52880, were taken in 1994 - the first two on 
HJD\,2449650.2 and the third on HD\,2449679.8.
}
\label{fig_c4}
\end{figure}

\subsection{Blue wing strengthenings (moderate velocities) }
\label{blwng}

  To gain some insight into the cause of the equivalent width variations 
of the red He\,I lines exhibited in Fig.\ref{fig_heew}, we examined the 
shapes of the two lines as a function of phase.  Figure~\ref{fig_heprof} 
illustrates that the cause of the equivalent width maximum of 
$\lambda$5875 at $\phi$ = 0.9 on each night is a strengthening of the 
blue wing. In these plots we exhibit this fact by multiplying the depth 
of $\lambda$6678 on the blue wing (and also extreme red wing) by a 
factor of 1.4 to 1.8. Figure~\ref{fig_heprof} shows that blue wings of the 
$\lambda$6678 scaled by these factors indeed replicate the $\lambda$5876 
absorptions at this phase. We see this phenomenon over a total range of 
0.08--0.10 cycles centered on radial velocity maximum ($\phi$$\approx$0.91) 
on each of the three nights. In these various examples the limiting 
line-depth scaling factor appears to be about 2.0. Since this is also 
the $gf$ ratio of the lines, it is likely that these extra absorptions 
are due to the medium at negative velocities being optically thin to 
He\,I line radiation. An alternative possibility, that the excess 
arises from metastability of $\lambda$5875, is implausible because no
such excess absorption is present in this line during the distension
(low-density) phases at $\phi$ $\sim$ 0.5. 

  If temperature variations cause the changes in the blue wings of the
He\,I lines, there should be a similar response in the {\it inner blue
wing} of the C\,IV doublet members - the cores of the Si\,IV lines are
too broad to demonstrate this. Figure~\ref{fig_c4} implies that the 
absorption-scaling argument is likely to be valid, that is,
that an excess absorption at about --100 km~s$^{-1}$ of line center is
caused by an optically thin column at these shifted wavelength. 
(By ``excess," as for the
He\,I lines in the previous figure, we refer to the absorption in the
blue wing of the stronger $\lambda$1548 line relative to $\lambda$1550.)
Massa's (1994; see Fig.\,3) grayscales exhibit this same effect -- as
a low-velocity ``spur" occurring $\sim$0.1 cycles before the wind
acceleration manifests itself at more negative velocities.

\subsection{C\,II line strength variations }

   The reader will recall that Young, Furenlid, \& Snow (1981) first
drew attention to C\,II variations during the cycle of BW\,Vul and 
posited that the C\,II variations were the result of the lines growing
anomalously strong outside the shock phases.
To test this assertion, we have synthesized the C\,II doublet with the
Hubeny {\it SYNSPEC} (Hubeny, Lanz \& Jeffery 1994) line synthesis 
code using Kurucz (static!) model atmospheres. For an 
atmosphere having $T_{\rm eff} = 23 000$K, $\log g = 4$, and $\xi_{\rm
  t} = 5$
km\,s$^{-1}$, we found an equivalent width of 0.22\,\AA\, for $\lambda$6578
and a $\lambda$6578/$\lambda$6583 ratio of 1.14. Nearly identical results 
obtain for $\log g = 3$. These values compare well with our mean {\it observed} 
$\lambda$6578 equivalent width value of 0.235$\pm{0.015}$\,\AA\, for phases 
outside the shocks and with the corresponding mean observed ratio of 1.19. 
Our modeling also shows that the strengths of these lines are quite 
insensitive to changes in temperature in the domain 20\,000--25\,000\,K.
The models do not confirm the speculation by YSF that changes in 
atmospheric density play a strong role in determining the line strength 
changes during the cycle. Of course, the line strengths will increase 
appreciably with any microturbulence that might accompany the shocks, 
so this would not explain the decrease during these phases.  
The upshot of these calculations is that the decreases in C\,II strengths
observed at shock phases cannot be due to simple changes in atmospheric
temperature or density. Thus, our results suggest that the question to pose
is not ``why are the C\,II strengths large during the non-shock phases?," 
but rather 
``{\it why do the C\,II strengths decrease to smaller values than they 
should have during shock phases?}" The question is critical
to a discussion of the shocks in BW Vul's atmosphere because the effects 
of temperature increases are obvious for other lines but lead to a 
contradiction for the response of the C\,II features. Adding additional 
turbulence from a shock would make the disagreement worse by forcing 
the predicted absorptions to be larger.

\begin{figure}
\includegraphics[width=84mm]{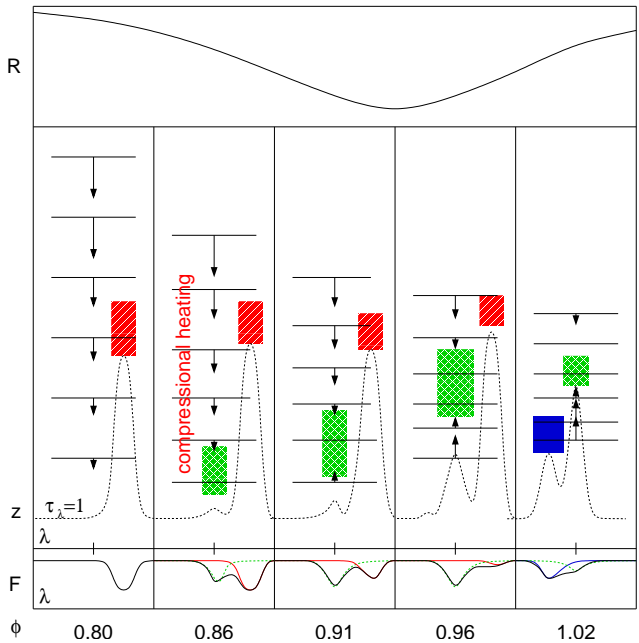}
\caption{Schematic illustration of the formation of the He\,I $\lambda$5876
 line profile as it evolves through radius minimum. The top panel
suggests the overall behaviour of the stellar radius. The five
bottom panels approximate the apparent line profile ($F(\lambda)$)
at five specific phases ($\phi$). The center of each panel corresponds
to the rest wavelength of the line (tick marks). 
The composite line profiles (black solid) are decomposed into 
red-shifted (red), blue-shifted (blue)
optically thick components and the unshifted optically thin component
(dotted green). The five vertical panels in the
center illustrate (i) the relative positions ($z$: horizontal lines) and
(ii) motions (vectors) of six specific Lagrangian zones in the atmospheres,
(iii) the position relative to these layers where the monochromatic
optical depth $\tau(\lambda)\approx1$ (black dotted line), and
(iv) the location where each component of line absorption is likely to
be strongest (shaded rectangles). 
Thus, for $\phi=0.86$, the optically thin stationary 
component ($\times$ shading) is formed {\it below} the optically thick red-shifted
 component ( / shading).
These profiles should be compared with those in Fig.~\ref{fig_heprof}, 
and also with Fig. 1 of Matthias et al. (1998).
}
\label{fig_cartoon}
\end{figure}

\subsection{Physical effect of the shocks on the He\,I and C\,IV lines }
\label{nub}

  If enhancements of neither temperature nor turbulence can explain 
the line weakening, let us consider instead the effect of a flattening 
of the temperature gradient with atmospheric depth, particularly in 
the construct of an LTE Milne-Eddington atmosphere. In this
formulation the strengths of weak lines should simply
be proportional to the gradient $dT/d\tau$. To interpret the enhanced 
absorptions in the blue wings of the He\,I lines (and possibly C\,IV) 
at $\phi$ = 0.9, one should also consider the heating effects from infall 
of material above where a line is being formed.
At this phase, upper strata are returning toward the surface at nearly 
free-fall velocities (Furenlid et al. 1987). Strata just below them fall 
slightly more slowly, and so on, down to the fully-braked stationary layers.
These combined decelerations produce compression over a broad range
of layers. The result of this pile up is a conversion of differential
flow velocity to local heating. Indeed, this effect can be observed 
as the first maximum of the UVC and He\,II line strength curves at $\phi$
= 0.85--0.1 (i.e., at velocity maximum) and by the {\it disappearance} of 
the cool-gas diagnostic, Si\,II $\lambda$6371, at the same times.
Heating will increase the number of atoms in
excited states and thus should permit new optically thin absorption
to appear in the excited He\,I lines. This absorption is formed in a
still-falling column which is still high enough in the atmosphere to be 
optically thin to an external observer. Most of the initial column density
will be concentrated near the shock, so the lobe will appear at near-rest
velocities. As this pile up proceeds, the threshold column density needed 
for visibility will recede (moving upward, in Eulerian coordinates),
permitting the optically thin absorption in the profile to grow toward 
positive velocities until it runs into and merges with the optically thick 
red lobe. The process ceases when the deepest layers essentially at rest 
become optically thick to the observer. At this point, just after phase 0.0, 
the entire profile becomes optically thick over a broad distribution of 
line velocities, both from the still-falling column and the strata at rest 
at the bottom. The standstill phase ends quickly as the material in the 
falling column is suddenly exhausted -- the optical depth turns thin and
to zero very quickly at wavelenths in the red lobe of the profile.

  During the infall phase the velocity gradient can be expected to 
increase among the superficial, cooler layers, causing proportionally 
more heating there. Thus, the weakening of the C\,II doublet (first
dip in Fig.~\ref{fig_c2ew}) is consistent with a flattening
of the gradient of the atmospheric temperature and line source 
functions.\footnote{A fact which may bear on this discussion is that to
reproduce UV spectrophotometic signatures from line aggregates in BW\,Vul's 
{\it IUE/SWP-camera} spectra, Smith (2001) found it necessary to impose
arbitrarily an artificial steepening of the temperature gradient for 
a simulated model atmosphere at minimum light phase. This is 
equivalent to imposing a flattening of the gradient in the maximum
light phase, as suggested from this optical study.} 
Indeed, in the expectation of accompanying increased turbulence, it seems 
difficult to understand how the weakening could arise in any other way. 
The second dip of the C\,II curve features coinciding with the passage of 
the pulsation shock can be explained by a more fundamental characteristic: 
the shock has a tendency to be more nearly isothermal than in the 
preshock atmosphere.

   The above picture is sketched qualitatively in Figure~\ref{fig_cartoon}. 
This illustration 
depicts the evolution of the stellar radius (top)
and the absorption profiles (bottom) during five phases ($\phi$), 
including one just 
prior to and one just following the primary shock passage interval. 
These line shapes may be compared to those in Fig.~\ref{fig_heprof} or 
Fig.~1 of Mathias et al. (1998) and
are additionally indicated as having blue-shifted, red-shifted and/or
stationary components. The five central panels show a representation
of several notional ``layers''. From left to right, these are
shown to be falling inwards (arrows), and then arrested and reversed by
material moving  upwards from beneath. Compression
gives rise to heating and a flattening of the temperature gradient. In
turn this gives rise to absorption in optically thin layers ($\phi=0.86$),
represented by a shaded rectangle ($\times$). Other line components
are also represented, displaced according to whether they are
formed in outflowing  (blue-shifted, solid), stationary 
(no shift, shaded $\times$ ) or infalling (red-shifted, shaded / ) material.
The zero velocity reference position or, equivalently, the rest wavelength 
for the absorption line is marked at the bottom of each panel. On the
basis of observations presented here, a
broken line indicates where the atmosphere becomes optically
thick  across the line profile ($\tau_{\lambda}=1$). Whether absorption
lines are formed in optically thick or thin layers is suggested by
their proximity to this curve. 
In any case, while 
admitting the omission of critical details of transfer of photons across 
the profile as the line as it turns optically thick, it appears possible 
that we can understand the formation of the excess optically thin 
absorption in the blue wing of the He\,I lines as well as the weakening 
of the temperature-sensitive C\,II and Si\,II lines. 
 
  Suffice it to say that in contrast to the infall shock the heating 
associated with the infall shock in our picture will extend over a 
broader range of strata in the atmosphere at any given moment than 
heating from the primary shock. If the UVC and He\,II and Si\,II line 
curves are to be taken as diagnostics of atmospheric heating, the 
pulsation shock is more impulsive and liberates more heat per unit 
time than the infall shock. In contrast, according to the equivalent 
width curves of Si\,II $\lambda$6371, the heating from the infall shock
not only lasts longer, but by differing amounts from cycle to cycle. 
This suggests that the details of this heating are driven by the 
strength of the earlier pulsation wave. Also, as noted in $\S$\ref{hilo}, 
the velocity jump criterion leads to the opposite 
inequality of apparent shock strengths, with the radial velocity jump
to the standstill (infall shock) being typically larger than the jump 
at the end of the standstill (primary shock). Since the velocity 
and equivalent width variations measure different types of shock 
properties, this apparent disagreement of the inequalities need not 
be a contradiction to the model.

\section{Conclusions}

   In this paper we have emphasized the importance of shock heating 
accompanying the passage of both shocks through the formation regions of 
several lines having a large range of excitation potentials (0--41 eV).
We have also pointed out by contrast that density variations are likely
to play at most a minor role in the observed line strength changes.
Several studies have remarked in the past on some of the optical and UV 
signatures we have discussed in detail, but the variations in the past
were attributed to kinematical, density, and/or finite propagation velocity 
effects -- rarely heating. Even the Young et al. (1987) study did not pursue
the subject beyond noting that variations of temperature-sensitive lines. 
This general inattention to the subject was not made through an explicit 
error by any of these investigators. In fact, our own assessment of the 
importance of temperature effects could not have come about except through 
echelle observations over a broad wavelength range, particularly including 
the red He\,I lines. Even so, the observation that the He\,I lines show 
strength variations due to atmospheric heating cannot be the whole story.  
Indeed, the realization that the C\,II doublet weakens during both shocks, 
and that the line strengths are insensitive to temperature variations, 
has all but forced the idea that the atmospheric temperature gradient is 
likely to be substantially flattened during shock passages. The heating of
the atmosphere causes the column length of excited atoms to increase, but
at the same time it causes the line source function gradient to decrease
slowly outwards, thereby producing line weakening of the saturated lines 
especially, such as the C\,II doublet.

   With this rough sketch in mind, we believe the time is ripe for new 
investigations of BW\,Vul to undertake a more quantitative analysis. This 
can begin with the abandonment of a static model atmosphere treatment, 
at least for the phase intervals in which shocks affect the atmosphere's 
thermal and density structure. 
Another logical step is to construct ``toy atmospheres" in which
radiative and/or hydrostatic equilibrium are abandoned to see 
what devices ``work," and to give some insight into the construction 
of a new generation of hydrodynamic models of shocks.  A line
synthesis from such kinematical atmospheres, in the spirit of Stamford \&
Watson's early modeling, can point the way to a more correct treatment 
of line doubling/broadening and strenth variations. One aspect especially
worth investigating is the effect of desaturation in various lines. 
One needs to know just how much pile-up material above the photosphere 
is required to give each type of lines a blue lobe strength that is 
some chosen fraction (say, one third) of the strength of the red lobe. 
In principle, combining these column densities with the observed rate of 
pile up at a given level will permit a test of our hypothesis that the 
Van Hoof effect in this star's atmosphere is an artifact of the optical 
thinesses of the blue lobes of these lines.

  In this paper the role of non-LTE has not been discussed. However,  
transfer of photons through the line profile is ultimately essential 
to the testing of our ideas of how the excess blue wing can develop
in the He\,I lines during the double-lobed infall-shock phase ($\phi$ = 0.9).
The details of line transfer and amount of turbulence
associated with the shocks are interrelated, and together they are critical
to a description of these lines at these phases. Another aspect of
non-LTE formation processes concerns the behavior of the He\,I lines 
during the low-density (maximum radius) phase. The equivalent widths
of the triplet $\lambda$5875 line give no indication that metastability
is important in the lower level (2$^{3}$P) of this transition, as would 
be expected for low (supergiant-like) atmospheric densities. This 
fact certainly means that the densities at this phase are not ``too 
low," but what this means quantitatively needs to be demonstrated. 
An even more sensitive indicator of low densities is the He\,I 
$\lambda$10830 line. Because of the importance of stimulated emission 
for infrared transitions, this feature is well known to exhibit a
strong response to either ``excess" absorption or emission.

   The heating of the base and rapid-acceleration regions of the wind also
remains to be evaluated. How hot these regions can become, and thus how
much pulsational energy is dissipated in the atmosphere, can be addressed 
by far-UV ({\it FUSE} satellite) observations of the O\,VI resonance doublet.

  Finally, one wonders about the effects of shocks in the outer atmospheres 
of other $\beta$\,Cep variables. Campos \& Smith (1981) and Smith (1983) 
examined high-resolution line profiles of several $\beta$\,Cep stars and 
found signatures of velocity jumps they associated with shocks in either 
optical or UV lines of all five of the ones for which line profiles they 
examined. In an analysis of $\nu$\,Eridani, Smith (1983) argued that rapid 
changes during key phases in the velocity curve and optical Si\,III line 
profiles of this star suggest the presence of atmospheric 
shocks, though not necessarily discontinuities in a radial velocity curve. 
Indeed, the equivalent width of the these lines grows probably due to the 
growth of a blue lobe. Unfortunately, other nonradial pulsation modes are 
likely to confuse the investigation of shock signatures. A more promising 
candidate may be another large-amplitude variable, $\sigma$\,Scorpii. 
Campos \& Smith (1980) have observed red-wing emission in the Si\,III 
triplet lines of this star near minimum velocity phase. Could the heating
associated with the pulsation shock in this star be occasionally so 
large as to produce emission? This star's line profiles have apparently 
not yet been extensively monitored since, but it might be profitable
to search for the cause of emission and to see how this relates to the
larger-amplitude shock of BW\,Vul for which no emissions are observed.

\section*{Acknowledgments}
   We wish to express our thanks to the McDonald Telescope Allocation 
Committee for their granting of three nights of 2.1-meter telescope time.
We are also grateful to Dr. Chris Johns-Krull for his reductions of the
McDonald data to echellogram format. We thank Dr. Jeff Valenti for the 
loan of his interactive wavelength calibration program and Mr. Anthony
Valenti for his time and instructions on the use of this program.

\label{lastpage}

\end{document}